\documentstyle[prl,aps,psfig]{revtex}
\newcommand{\beq}{\begin{equation}}
\newcommand{\eeq}{\end{equation}}
\newcommand{\beqa}{\begin{eqnarray}}
\newcommand{\eeqa}{\end{eqnarray}}
\newcommand{\ba}{\begin{array}}
\newcommand{\ea}{\end{array}}

\begin{document}
\draft

\twocolumn[\hsize\textwidth\columnwidth\hsize\csname
@twocolumnfalse\endcsname

\widetext 
\title{Modulational Instability and Complex Dynamics 
of Confined Matter-Wave Solitons} 
\author{L. Salasnich$^{1}$, A. Parola$^{2}$ and L. Reatto$^{1}$} 
\address{$^{1}$Istituto Nazionale per la Fisica della Materia, 
Unit\`a di Milano, 
Dipartimento di Fisica, Universit\`a di Milano, \\
Via Celoria 16, 20133 Milano, Italy\\
$^{2}$Istituto Nazionale per la Fisica della Materia, 
Unit\`a di Como, 
Dipartimento di Scienze Fisiche, Universit\`a dell'Insubria, \\
Via Valeggio 11, 23100 Como, Italy}

\maketitle

\begin{abstract} 
We study the formation of bright solitons in a Bose-Einstein 
condensate of $^7$Li atoms induced by a sudden change in the sign 
of the scattering length from positive to negative, 
as reported in a recent experiment (Nature {\bf 417}, 150 (2002)). 
The numerical simulations are performed by 
using the 3D Gross-Pitaevskii equation (GPE) with a 
dissipative three-body term. We show that a number of 
bright solitons is produced and this can be interpreted 
in terms of the modulational instability of the 
time-dependent macroscopic wave function of the Bose condensate. 
In particular, we derive a simple formula for the number 
of solitons that is in good agreement with the numerical results of 3D GPE. 
By investigating the long time 
evolution of the soliton train solving the 1D GPE with three-body 
dissipation we find that adjacent solitons repel each other 
due to their phase difference. 
In addition, we find that during the motion of 
the soliton train in an axial harmonic potential the number of 
solitonic peaks changes in time and the density of individual peaks 
shows an intermittent behavior. Such a complex dynamics explains the 
``missing solitons'' frequently found in the experiment. 
\end{abstract}

\pacs{03.75.Fi; 32.80.Pj; 42.50.Vk}

]

\narrowtext

The recent experimental observation of dark \cite{1} and 
bright \cite{2,3}  solitons in Bose-Einstein condensates 
has renewed the interest 
in the intriguing 
dynamical phenomena of nonlinear matter waves \cite{4}. 
In the experiment of Strecker {\it et al.} \cite{3} soliton trains have been 
formed by making a stable condensate of $^7$Li atoms with a 
large positive scattering length $a_s$ using a Feshbach resonance
and then switching $a_s$ to a negative value. 
The formation of the soliton train has been interpreted 
as due to quantum mechanical 
phase fluctuations of the Bosonic field operator \cite{5}. 
By imposing a suitable space dependent pattern in the 
initial phase of the Bose condensate 
and then using the 1D time-dependent 
Gross-Pitaevskii equation (GPE), Al Khawaja {\it et al.} \cite{5} 
have reproduced the formation of the soliton train. 
\par 
In this paper we show that 
the formation and subsequent evolution 
of a soliton train can be adequately investigated by using the 
classical (mean-field) time-dependent 3D GPE 
with a dissipative term which takes into account the 
three-body recombination process that is crucial 
during the collapse of the condensate \cite{6}. 
The multi-soliton configuration is obtained without 
imprinting the initial wave function with a fluctuating phase. 
We show that the soliton train 
is produced by the modulational instability (MI) of the evolving 
classical phase of the Bose condensate (see also \cite{7}). 
MI is a nonlinear wave phenomenon in which an exponential 
growth of small perturbations  
results from the interplay between nonlinearity and anomalous 
dispersion. MI has been previously 
studied for waves in fluids \cite{8}, in plasma physics \cite{9}, 
in nonlinear optics \cite{10}, and also in the context of the 
superfluid-insulator transition of a Bose-Einstein condensate 
trapped in a periodic potential \cite{11}. Here we find that the number of 
bright solitons induced by MI is given by a simple analytical 
formula which reproduces our numerical simulations. 
By investigating the long time evolution of the soliton train 
under the action of a harmonic potential of frequency $\omega_z$ 
in the travelling axial direction we find that the center of mass 
motion is periodic with frequency $\omega_z$ but the density 
of each solitonic peak strongly changes in time. 
\par 
At zero temperature the macroscopic wave function 
$\psi({\bf r},t)$ of a Bose-Einstein condensate made of $^7$Li 
atoms can be modeled by the following dissipative 
time-dependent Gross-Pitaevskii equation 
\beq 
\left[ 
i\hbar {\partial \over \partial t} 
+{\hbar^2\over 2 m} \nabla^2 - U  
- g N |\psi|^2 + i \gamma N^2 |\psi|^4 \right] \psi = 0  \; , 
\eeq 
where $g=4\pi\hbar^2 a_s/m$, $a_s$ is the s-wave scattering length, 
$m$ the atomic mass, $N$ is the number of condensed atoms and 
$\gamma$ is the strength of the dissipative three-body term \cite{6}. 
At $t=0$ the condensate wave function is normalized to one. 
Following the experiment of Strecker {\it et al.} \cite{3}, 
the external potential $U({\bf r})$ is given by 
\beq 
U({\bf r}) = {m\over 2} 
\left[\omega_{\bot}^2 (x^2+y^2) + {\chi^2} \omega_z^2 z^2 \right] + V_L(z) 
\eeq 
where $V_L(z)$ is the box optical potential that initially 
confines the condensate in the longitudinal direction. 
The harmonic confinement is anisotropic with 
$\omega_{\bot}=2\pi \times 800$ Hz and $\omega_z =2\pi \times 4$ Hz. 
$\chi$ is a parameter that modifies the intensity of 
confinement in the axial direction. 
\par 
To investigate the formation of the train of bright solitons, 
we first calculate the ground-state wave function of the condensate 
with a positive scattering length $a_s=100 a_B$, where $a_B=0.53\AA$ is 
the Bohr radius, by using Eq. (1) with imaginary time and 
$\gamma = 0$. The numerical code implements in cylindric symmetry $(z,r)$ 
a finite-difference splitting method (spatial grid 
with $400\times 100$ points) with a predictor-corrector 
algorithm to treat the nonlinear term \cite{12}. 
We choose a condensate with longitudinal width 
$L=284.4$ $\mu$m and $N=10^4$ atoms. 
Then we use the ground-state wave function 
as initial condition for the time-evolution 
of Eq. (1) with $a_s=-3a_B$, $V_L(z)=0$ and $\chi =0$. 
Following \cite{6} we choose $\gamma = 1.77\times 10^{-11}$ 
$(\hbar \omega_z)/a_z^6$. $\gamma$ is important during the 
collapse, because is fixes the critical density at which 
the compression ceases. 

\begin{figure}
\centerline{\psfig{file=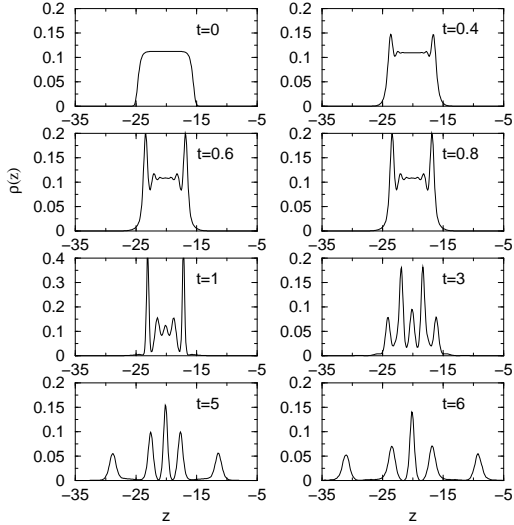,height=2.8in}} 
\caption{Axial density profile $\rho(z)$ of the 
Bose-Einstein condensate made of $10^4$ $^7$Li atoms 
obtained by solving dissipative 3D GPE. 
For $t<0$ the scattering length is $a_s=100a_B$, while 
for $t\geq 0$ we set $a_s=-3a_B$ with $a_B$ the Bohr radius 
and $V_L=0$. External harmonic potential given by 
Eq. (2) with $\chi=0$. 
Length $z$ in units $a_{z}=(\hbar /m\omega_z)^{1/2}$, 
density $\rho$ in units $1/a_z$ and 
time $t$ in units $\omega_z^{-1}$. } 
\end{figure} 

In Figure 1 we plot the probability density 
in the longitudinal direction $\rho(z)=\int dxdy |\psi(x,y,z)|^2$ 
of the evolving wave-function. 
The initially homogeneous condensate shows the formation of 5 peaks. 
Note that initially the phase of the condensate is set equal to zero. 
After the formation, the peaks start to separate each other showing a 
repulsive force between them. Eventually each peak evolves 
with small shape oscillations but without appreciable dispersion. 
This means that strictly speaking the train is not made 
of shape-invariant waves.  
We will use the word ``solitons'' to indicate the peaks of the train 
because they remain spatially localized during the time evolution. 
The formation of these solitons can be explained 
as due to the modulational instability 
(MI) of the time-dependent wave function of the Bose condensate, 
driven by imaginary Bogoliubov excitations \cite{5}. 
The Bogoliubov elementary excitations $\epsilon_k$  
of the static Bose condensate $\Phi({\bf r})$ 
can be found from Eq. (1) by looking for solutions of the form 
$
\psi({\bf r},t) = e^{-i\mu t/\hbar}
\left[ \Phi({\bf r}) + u_k({\bf r}) e^{-i\epsilon_k t/\hbar}
+ v_k^*({\bf r}) e^{i\epsilon_k t/\hbar}
\right] , 
$
and keeping terms linear in the complex functions 
$u({\bf r})$ and $v({\bf r})$. Neglecting the dissipative 
term $\gamma$, in the quasi-1D limit [13] one finds 
$
\epsilon_k = \sqrt{ {\hbar^2 k^2/(2m)} 
\left({\hbar^2 k^2/(2 m)} 
+ 2 g_{1D} n \right) }  , 
$ 
where $g_{1D}=g/(2\pi a_{\bot}^2)$ with 
$a_{\bot}=\sqrt{\hbar/(m\omega_{\bot})}$, $n=N/L$ is the 
linear density and $L$ is the length of the condensate. 
By suddenly changing the scattering length $a_s$ to a negative value, 
the excitations frequencies corresponding to 
$k<k_c = \sqrt{16 \pi |g_{1D}| n}$ become 
imaginary and, as a result, small perturbations grow 
exponentially in time. 
It is easy to find that the maximum rate of growth is at 
$k_0=k_c/\sqrt{2}$. The wavelength of this mode is 
$\lambda_0 = {2\pi / k_0}$ and the ratio $L/\lambda_0$ gives 
an estimate of the number $N_s$ of bright solitons which 
are generated: 
$
N_s = {\sqrt{N|a_s|L}/(\pi a_{\bot}) } . 
$

\begin{figure}
\centerline{\psfig{file=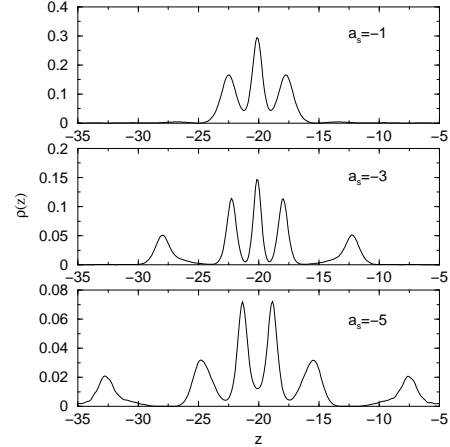,height=2.4in}}
\caption{Axial density profile $\rho(z)$ of the 
Bose-Einstein condensate made of $10^4$ $^7$Li atoms 
at $t=4.6$ obtained by solving dissipative 3D GPE for three values 
of the final scattering length $a_s$. 
External harmonic potential given by 
Eq. (2) with $\chi=0$. 
Scattering length $a_s$ in units 
of the Bohr radius $a_B$. Units as in Figure 1.} 
\end{figure} 

As predicted by this formula, Figure 2 shows that the number 
of peaks grows with the scattering length. The predicted number 
$N_s$ of solitons is in very good agreement with the 
numerical results: $N_s=2.60$ for $a_s=-1 a_B$, $N_s=4.51$ for 
$a_s=-3a_B$ and $N_s=5.82$ for $a_s=-5a_B$.   
The period $\tau_0 = \hbar /Im\left({\tilde \epsilon}_{k_0}\right)$ 
associated to the most unstable mode is given by 
$ 
\tau_0 = {mL^2 / (\pi \hbar N_s^2) } ,   
$ 
in rough agreement with the numerically estimated 
formation time of the strongest fluctuations in the number 
of peaks belonging to the soliton train. 
\par 
In the experiment of Strecher {\it et al.} \cite{3} there are 
initially about $10^5$ atoms. 
In this case our numerical (see below) and analytical 
results predict the formation of about $N_s=15$ solitons, 
that is twice 
the number experimentally detected. We have verified that 
the number of solitons does not depend on the 
dissipation constant $\gamma$ of Eq. (1) but increasing 
$\gamma$ the densities of solitons are reduced and 
their widths are increased. Note that in this system  
recombination processes have not been investigated 
experimentally and these phenomena can affect the imaging process 
of soliton train densities. 
\par 
We have also verified that the number of solitons increases 
by increasing $\Delta t$, the delay time between the removal of 
endcaps and the change of scattering length 
(in figure 1 and 2 it is $\Delta t = 0$). 
For instance, with the data and units of Figure 2, 
$N_s=6$ for $\Delta t = 0.3$ and $N_s=7$ for $\Delta t = 0.6$. 
That is a simple consequence of the enlargement 
of the axial width of the Bose-Einstein condensate, 
in full agreement with the experimental results \cite{3}.  

\begin{figure}
\centerline{\psfig{file=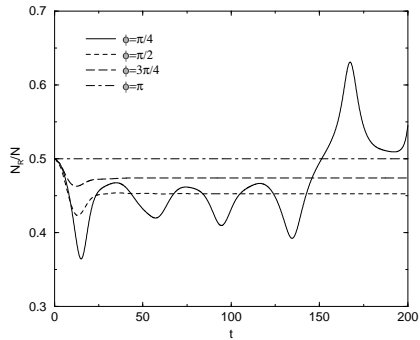,height=1.8in}}
\caption{Ratio $N_R/N$ as a function of time 
for two neighboring solitons with an initial 
phase difference $\phi$ and spatial separation $2z_0 = 6$.  
$N=N_R+N_L$ is the sum of the number $N_R$ of atoms 
in the right soliton and the number $N_L$ of atoms 
in the left soliton. $N|a_s|/a_{\bot}=1/\sqrt{10}$. 
Results obtained by solving conservative NPSE. 
External harmonic potential given by 
Eq. (2) with $\omega_{\bot}/\omega_z =10$ and 
$\chi=0$. Units as in Figure 1.} 
\end{figure}

\par 
In order to clarify the role of phases 
in the interaction between bright solitons 
we consider a simpler model: two solitons 
without dissipation. 
It has been found that the effective interaction between two 
bosonic matter waves depends on their phase difference $\phi$ , 
being proportional to $a_s\cos{(\phi)}$ \cite{5,14}. 
The axial wave function $f(z)$ of a bright soliton 
under transverse harmonic confinement can be analytically 
determined \cite{14} by using the nonpolynomial Schr\"odinger 
equation (NPSE), an effective 1D equation 
we have recently derived from the 3D GPE \cite{15}. 
In \cite{14,15} we have shown that 1D GPE reproduces 
bright solitons of 3D GPE only when the Bose condensate 
is strongly cigar-shaped. Instead, NPSE always reproduces 
3D GPE solitons with great accuracy. 
As initial condition for the numerical solution of the 
time-dependent NPSE (spatial grid of $10^4$ points) 
we choose 
$\psi(z)=\left[f(z-z_0)+f(z+z_0)e^{i\phi}\right]/\sqrt{2}$, 
where $\phi$ is the phase difference of the two neighboring solitons 
centered in $-z_0$ and $z_0$. 
\par
In Figure 3 we plot the time evolution of $N_R/N$, 
where $N_R$ is the number of atoms in the 
right soliton, for some values of $\phi$ 
with $N|a_s|/a_{\bot}=1/\sqrt{10}$, $\omega_{\bot}/\omega_z=10$ 
and $\chi=0$. For $\phi = 0$ the two solitons are attractive and 
eventually form a static peak which radiates small waves. 
For $\phi = \pi/4$ the centers of the two solitons 
do not change with time but they exchange atoms 
as in a Josephson junction. Obviously the amount of atom 
exchange will depend on the details of the two solitons 
at initial time (widths and separation distance). 
We find 
such a complex exchange of atoms for $0<\phi<\pi/2$, 
while for $\phi = \pi/2$ the two solitons 
are weakly repulsive, their shapes change with time and 
eventually two solitons of different density appear 
because the Josephson exchange cannot continue when 
the two solitons separate due to their repulsive 
interaction. 
Note that with $\phi=-\pi/2$ the densities of the two 
peaks are inverted. For larger values of $\phi$ 
the dynamics is quite similar to the 
$\phi=\pi/2$ case, with two solitons of different shapes 
getting away from each other. The fraction of 
exchanged atoms between the two solitons 
goes to zero as $\phi$ approaches $\pi$  
due to the increase of repulsive interaction and 
the decrease of interaction time. 
Exactly at $\phi =\pi$ parity symmetry of the problem 
inhibit atom exchange. 

\begin{figure*}
\centerline{\psfig{file=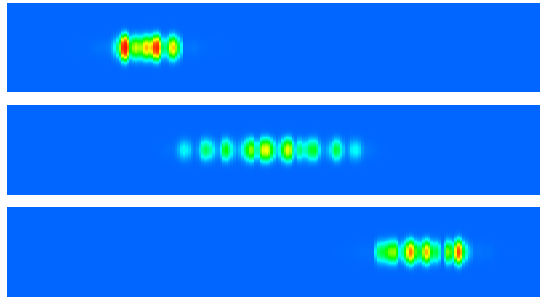,height=1.7in}}
\caption{Moving soliton train made of 
Bose condensed $^7$Li atoms 
obtained by solving the dissipative 1D GPE. Initially 
there are $N=10^5$ atoms. Three frames: 
$t=0.8$ (top), $t=1.6$ (middle), and $t=3.8$ (bottom). 
External harmonic potential given by 
Eq. (2) with $\chi=1$. Red color corresponds to 
highest density. Units as in Figure 1.} 
\end{figure*}

\par
In the soliton train shown in Figure 1 and Figure 2 
the phase difference of adjacent solitons is 
not imposed a priori at $t=0$, like in Ref. \cite{5}, 
but is self consistently determined by the GPE. 
As a consequence, its value is not 
exactly equal to $\pi$ and it changes with
time. It is interesting to study the long time evolution 
of the soliton train under axial harmonic confinement 
as done in the experiment of Strecker {\it et al.} 
using $10^5$ $^7$Li atoms \cite{3}. 
Instead of 3D GPE in this numerical simulation 
we use the 1D GPE (spatial grid of $10^4$ points) 
with rescaled effective interaction 
and dissipative term \cite{15}, which allows to extend our 
calculations to long times. As explained in \cite{15}, 
1D GPE is obtained from 3D GPE by imposing a Gaussian 
wave function of width $a_{\bot}$ in the transverse direction. 
Note that we have verified 
that the 1D GPE results are in good agreement 
with 3D GPE ones for the set of parameters of Streker 
{\it et al.} experiment \cite{3}. 
\par
In Figure 4 we show three frames of the traveling train as 
color contour-plots of the density. In each pannel the orizontal axis 
is the $z$ coordinate and the vertical axis is the $x$ coordinate. 
The center of mass of the soliton train oscillates with 
the frequency of the 
harmonic confinement. Moreover, the solitons spread out in the middle 
and bunch at the turning points in very good agreement with 
the experimental results under similar conditions \cite{3}. 
\par
It is interesting to observe that during 
the time evolution the densities of bright solitons 
oscillate in an irregular way and, at certain instants, 
a few solitons practically disappear and reappear 
after a while. Such intermittent behavior is related 
to the complex dynamics of the bright soliton phases due to 
the presence of axial harmonic confinement. 
As confirmed in Figure 5, in absence of axial 
confinement ($\chi=0$), after a transient 
the peaks, separating each other, become true solitons. 
In particular, the number $N_p$ of peaks in the train 
gets constant and equal to the analytically 
estimated number $N_s$ of bright solitons. 
Instead, in presence of axial harmonic confinement ($\chi = 1$),  
$N_p$ changes in time. 
Obviously, in this case the peaks never become true solitons. 

\begin{figure}
\centerline{\psfig{file=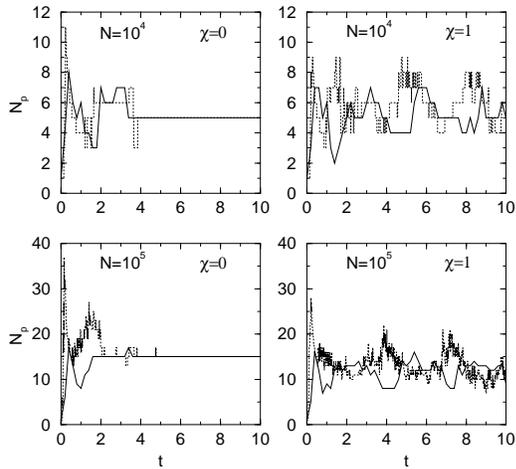,height=2.5in}}
\caption{Number $N_p$ of peaks in the 
train of Bose condensed $^7$Li atoms as a function of time $t$. 
$N$ is the initial number of atoms. 
Simulation performed by using the dissipative 1D GPE (dotted line). 
The solid line is obtained from a smoothed finite-resolution 
density profile of the train. 
External harmonic potential given by Eq. (2). 
Units as in Figure 1.} 
\end{figure}
\par 
In the experiment of Strecker {\it et al.} \cite{3}  
``missing solitons'' are 
frequently observed during the time evolution of the train 
\cite{16}. Our results strongly suggest that 
the phenomenon of ``missing solitons'' is related 
to the intermittent 
dynamics of individual peaks in the train under axial 
harmonic confinement. 
\par
To simulate the finite resolution of the experimental detection 
and imaging process, we calculate the covolution 
${\bar \rho}(z)=\int d{z'} G(z-{z'})\rho({z'})$ of the axial density 
profile $\rho(z)$ with a Gaussian $G(z)$ having the width of 
a single soliton. As shown in Figure 5, for $\chi =1$ 
the number $N_p$ of peaks of the smoothed 
density ${\bar \rho}(z)$ oscillates in time around a mean value 
smaller than $N_s$. In fact, many peaks of the train aro too close  
to be seen by using a finite resolution of the density:  
that becomes dramatic at the turning points. 
Thus, the effect of finite resolution could explain 
the disagreement between the experimentally 
observed number of solitons and its analytical estimate 
based on modulational instability. 
\par 
In Figure 5 it is also shown the behavior of $N_p$ 
with initially $10^4$ atoms. Remarkably, in this case 
the largest values of $N_p$ are obtained when the 
train is in the middle of the trap: contrary to the case 
with initially $10^5$ atoms, the solitons 
bunch in the middle and spread out at the turning points. 
Such a phenomenon can be qualitatively explained observing 
that the energy of the repulsive interaction between solitons 
decreases by reducing the number of atoms \cite{5,14}. 
It follows that below a critical threshold in the number of atoms 
the repulsive interaction between solitons is overcome by the 
potential energy of solitons and the solitons cross 
in the middle of the trap.  
We have verified that to get spreading solitons in the middle 
of the trap with initially $10^4$ atoms it is necessary 
to strongly reduce the axial frequency ($\chi = 1/4$). 
\par
In conclusion, we have successfully explained and numerically 
simulated the dynamical process of soliton train formation induced 
by modulational instability in the framework of the 
time-dependent Gross-Pitaevskii equation with a three-body 
dissipative term. Contrary to the claim of Al Khawaja {\it et al.} 
\cite{5}, it is not necessary to include quantum phase fluctuations to 
trigger the formation of the soliton train. 
We have also investigated the 
soliton-soliton interaction and found a novel phenomenon: 
the intermittent dynamics of individual peaks during the 
time evolution of the soliton train in an axial harmonic potential. 
Signatures of this phenomenon can be extracted from the data 
of the experiment of Strecker {\it et al.} \cite{3}. 
Because of the intimate connection between atom optics 
with Bose-Einstein condensates and light optics we believe that 
the intermittent dynamics in soliton trains may be also 
observed with optical solitons in fibres. 
\par 
L.S. thanks R.G. Hulet for many useful e-discussions.


\begin{thebibliography}{99}
\bibitem{16} It is important to stress that in the experiment 
the imaging process is destructive: to follow the time evolution 
of the train the authors of \cite{3} had to repeat many 
times the experiment with the same initial conditions.
\end{thebibliography}
\end{document}